\documentclass[aps, showpacs,showkeys,pra] {revtex4}
\usepackage{hyperref}
\usepackage{amsmath }
\usepackage{amssymb}
\usepackage{epsfig}
\usepackage{natbib}
\newcommand*{\be}{\begin{equation}}
\newcommand*{\ee}{\end{equation}}

\begin{document}
\bibliographystyle{revtex}
\title{ Rashba coupling in quantum dots in the presence of magnetic field}
\author{V.V. Kudryashov}
 \email{kudryash@dragon.bas-net.by}
\affiliation{Institute of Physics, National Academy of Sciences of
Belarus \\68 Nezavisimosti  Ave., 220072, Minsk,  Belarus }

\begin{abstract}{We present an analytical solution to the Schr\"odinger equation for  electron in
a two-dimensional circular quantum dot in the presence of both external magnetic field and the
 Rashba spin-orbit interaction. The confinement is described by  the realistic potential well of
 finite depth.}
\end{abstract}

\maketitle

\section{Introduction}

The Schr\"odinger equation describing  electron in a two-dimensional quantum dot normal to the $z$ axis is of the form
\begin{equation}
 \left(\frac{{\bf P}^2}{2 M_{eff}} + V_{c}(x,y) +V_R + V_Z \right) \Psi = E
 \Psi ,
\end{equation}
where $M_{eff}$ is the effective electron mass. The vector potential
${\bf A} =\frac{B}{2}(-y,x,0)$ of a magnetic field oriented perpendicular to the plane of the quantum dot leads to the generalized momentum ${\bf P} ={\bf p} + \frac{e}{c}{\bf A}.$
We have the usual expression for the Zeeman interaction
\begin{equation}
V_Z =\frac{1}{2}g \mu_B B\sigma_z ,
\end{equation}
where $g$ represents the effective gyromagnetic factor, $\mu_B$ is the Bohr's magneton
The Rashba
spin-orbit interaction  \cite{ras,byc} is represented as
\begin{equation}
V_R= a_R (\sigma_x P_y - \sigma_y P_x) .
\end{equation}
The  Pauli spin-matrices are defined as standard,
\[
  \sigma_x = \left(\begin{array}{cc}
 0&1 \\
 1&0
 \end{array}\right), \quad
\sigma_y =  \left(\begin{array}{cr}
 0&-i \\
 i&0
 \end{array}\right), \quad
   \sigma_z = \left(\begin{array}{cr}
 1&0 \\
 0&-1
 \end{array}\right).
\]

A confining potential is usually assumed to be
symmetric,$V_c(x,y)= V_c(\rho), \rho =\sqrt{x^2 +y^2}$.
There are two model potentials which are widely employed in this area.
The first is a harmonic oscillator potential  \cite{val,kua}. Such a model  admits the approximate
(not exact) solutions of Eq. (1). The second model is a circular quantum dot with hard walls \cite{bul,tsi}
 $ V_{c}(\rho)=0$ for $\rho <  \rho_0$, $
V_{c}(\rho)= \infty$ for $\rho >  \rho_0$ .
 This model is  exactly solvable. In the framework of above models the number
of allowed energy levels   is infinite for the fixed  total angular momentum in the
absence of a magnetic field.

   In this paper, we propose new model  which corresponds to a circular quantum dot with a  potential well of finite depth:  $V_c(\rho) =0$ for $\rho <  \rho_0$, $
V_{c}(\rho)= V = constant$ for $\rho >  \rho_0$ . Our model is exactly solvable and the number
of admissible energy levels is  finite for the fixed  total angular momentum in the absence of a magnetic field. The present solutions contain, as limiting cases, our previous results \cite{kud} (no external magnetic field).

\section{Analytical solutions of the  Schr\"odinger equation}

 The Schr\"odinger equation (1)
 is considered  in the cylindrical coordinates
 $
 x = \rho \cos \varphi,  y = \rho \sin \varphi .
 $
 Further it
is convenient to employ dimensionless  quantities
 \begin{equation}
r= \frac{\rho}{ \rho_0}, \quad \epsilon =\frac{2 M_{eff}}{\hbar^2} \rho_0^2 E, \quad
  v= \frac{2 M_{eff}}{\hbar^2} \rho_0^2 V , \quad
  a=\frac{2 M_{eff}}{\hbar} \rho_0 a_R , \quad b = \frac{e B \rho_0^2}{2 c \hbar}, \quad s = \frac{g M_{eff}}{4 M_e} .
 \end{equation}
 Here $M_e$ is the electron mass.
As it was shown in \cite{bul} equation (1) permits the separation of variables
\begin{equation}
 \Psi_m(r,\varphi) =  u(r) e^{i m \varphi}\left(\begin{array}{c}
 1 \\
  0
 \end{array}\right) + w(r)  e^{i (m+1) \varphi} \left(\begin{array}{c}
 0 \\
  1
 \end{array}\right) ,\quad m=0, \pm 1, \pm 2, \ldots
\end{equation}
due to conservation of the total angular momentum
$L_z + \frac{\hbar}{2} \sigma_z $.

We have the following radial equations
\begin{eqnarray}
 \frac{d^2u}{dr^2} + \frac{1}{r} \frac{d u}{d r}  +(\epsilon -v)u - \frac{m^2}{r^2}u -2 b m u - b^2 r^2 u
 - 4 s b u \nonumber \\
= a  \left(\frac{d w}{d r} +  \frac{m+1}{r} w  + b r w\right) ,
\nonumber \\
  \frac{d^2w}{dr^2} + \frac{1}{r} \frac{d w}{d r}  +(\epsilon -v)w - \frac{(m+1)^2}{r^2}w -2 b (m+1) w - b^2 r^2 w
 + 4 s b w \nonumber \\
= a  \left(-\frac{d u}{d r} +  \frac{m}{r}u  + b r u\right) .
\end{eqnarray}
In \cite{bul,tsi}, the requirements $ u(1)= w(1)=0 $
 were imposed. In our model, we look for the radial wave functions
$u(r)$ and $w(r)$ regular at  the origin $r=0$ and decreasing at
infinity $r \rightarrow \infty$.

Following \cite{tsi} we use the substitutions
\begin{equation}
u(r) = \exp\left(\frac{-b r^2}{2}\right) (\sqrt{b} r)^{|m|}  f(r),
\quad w(r) = \exp\left(\frac{-b r^2}{2}\right) (\sqrt{b}
r)^{|m+1|} g(r)
\end{equation}
which lead to the confluent hypergeometric equations in the case $a=0$. Therefore we attempt to express the desired
solutions  of Eq. (6) via the confluent hypergeometric functions when $a \neq 0$.

We consider two regions $ r <1$ (region 1) and $r >1$  (region 2)
separately.

 In  the region 1  ($ v = 0$), using the known properties
 \begin{eqnarray}
 M(\alpha,\beta,\xi) - \frac{d M(\alpha,\beta,\xi)}{d \xi} = \frac{\beta -\alpha}{\beta} M(\alpha,\beta + 1,\xi) ,
 \nonumber \\
 (\beta - 1  -\xi)M(\alpha,\beta,\xi) +\xi \frac{d M(\alpha,\beta,\xi)}{d \xi} = (\beta - 1)M(\alpha -1,\beta -1,\xi)
 \end{eqnarray}
of the confluent hypergeometric functions $ M(\alpha,\beta,\xi)$ of the first kind  \cite{abr} it is easily to
show that
 the suitable particular solutions of the radial equations  are
\begin{eqnarray}
u_1(r) = \exp\left(\frac{-b r^2}{2}\right) (\sqrt{b} r)^{|m|}
\left(c_{1-} f_{1-}(r) + c_{1+} f_{1+}(r) \right),
\nonumber \\
 w_1(r) = \exp\left(\frac{-b r^2}{2}\right) (\sqrt{b} r)^{|m+1|}\left(\frac{a }{2 \sqrt{b}}\right) \left(c_{1-} g_{1-}(r) + c_{1+} g_{1+}(r) \right),
\end{eqnarray}
 where
\begin{eqnarray}
 f_{1\mp}(r) = M(m+1- k^{\mp}_1,m+1,b r^2),
\nonumber \\
g_{1\mp}(r) = \left(\frac{ k^{\mp}_1}{(m+1)}\right) \frac{M(m+1- k^{\mp}_1,m+2,b r^2)}{(-k^{\mp}_1 + (4 b)^{-1} \epsilon +s - 1/2)}
\end{eqnarray}
for $m=0,1,2...$,
\begin{eqnarray}
 f_{1\mp}(r) = M(1- k^{\mp}_1,-m+1,b r^2),
\nonumber \\
g_{1\mp}(r) = m \frac{M(- k^{\mp}_1,-m,b r^2)}{(-k^{\mp}_1 + (4
b)^{-1} \epsilon +s - 1/2)}
\end{eqnarray}
for $m=-1,-2,-3...$  and
\begin{equation}
k^{\pm}_1 = \frac{1}{4 b} \left( \epsilon + \frac{a^2}{2} \pm a \sqrt{\epsilon +\frac{a^2}{4}
 + \left(\frac{4 b}{a}\right)^2 (s - 1/2)^2} \right) .
 \end{equation}
Here $c_{1-}$ and $c_{1+}$ are arbitrary coefficients.
The functions $u_1(r)$ and $w_1(r)$ have the desirable behavior at
the origin.

In  the region 2  ($ v > 0$), using the known properties
\begin{eqnarray}
U(\alpha,\beta,\xi) - \frac{d U(\alpha,\beta,\xi)}{d \xi} = U(\alpha,\beta + 1,\xi) ,
\nonumber \\
(\beta - 1 - \xi)U(\alpha,\beta,\xi) + \xi \frac{d U(\alpha,\beta,\xi)}{d \xi} = - U(\alpha -1,\beta - 1,\xi)
\end{eqnarray}
   of the confluent hypergeometric functions $ U(\alpha,\beta,\xi)$  of the second kind  \cite{abr} it is simply to get the suitable real solutions of the radial equations:
\begin{eqnarray}
u_2(r) = \exp\left(\frac{-b r^2}{2}\right) (\sqrt{b} r)^{|m|}
\left(c_{2-} f_{2-}(r) + c_{2+} f_{2+}(r) \right),
\nonumber \\
 w_2(r) = \exp\left(\frac{-b r^2}{2}\right) (\sqrt{b} r)^{|m+1|}\left(\frac{a }{2 \sqrt{b}}\right) \left(c_{2-} g_{2-}(r) + c_{2+} g_{2+}(r) \right),
\end{eqnarray}
where
\begin{eqnarray}
f_{2\mp}(r) = \frac{\sqrt{\mp 1}}{2}\left( U(m+1-k^-_2,m+1,b r^2)  \mp   U(m+1-k^+_2,m+1,b r^2)\right) ,
\nonumber \\
g_{2\mp}(r)=  \frac{\sqrt{\mp 1}}{2}\left( \frac{U(m +1- k^-_2,m+2,b r^2)}{(-k^-_2 + (4 b)^{-1} (\epsilon -v) +s - 1/2)} \mp
\frac{U(m+1- k^+_2,m+2,b r^2)}{(-k^+_2 + (4 b)^{-1} (\epsilon-v)+s - 1/2)} \right)
\end{eqnarray}
for $m=0,1,2...$,
\begin{eqnarray}
f_{2\mp}(r) = \frac{\sqrt{\mp 1}}{2}\left( U(1-k^-_2,-m+1,b r^2)  \mp   U(1-k^+_2,-m+1,b r^2)\right) ,
\nonumber \\
g_{2\mp}(r)=  \frac{\sqrt{\mp 1}}{2}\left( \frac{U(- k^-_2,-m,b r^2)}{(-k^-_2 + (4 b)^{-1} (\epsilon -v) +s - 1/2)} \mp
\frac{U(- k^+_2,-m,b r^2)}{(-k^+_2 + (4 b)^{-1} (\epsilon-v)+s - 1/2)} \right)
\end{eqnarray}
for $m=-1,-2,-3...$ and
\begin{equation}
k^{\pm}_2 = \frac{1}{4 b} \left( \epsilon - v + \frac{a^2}{2} \pm i a \sqrt{v -\epsilon -\frac{a^2}{4}
 - \left(\frac{4 b}{a}\right)^2 (s - 1/2)^2} \right) .
\end{equation}
Here $c_{2-}$ and $c_{2+}$ are arbitrary coefficients.
The functions $u_2(r)$ and $w_2(r)$ have the appropriate behavior at
infinity.

We assume the realization of condition
\begin{equation}
\epsilon < v* =v  -\frac{a^2}{4}
 - \left(\frac{4 b}{a}\right)^2 (s - 1/2)^2
\end{equation}
 which means that electron belongs to a quantum dot.  We can also obtain the exact solutions
when $\epsilon > v*$. However, in this case we cannot consider electron as belonging to
 a quantum dot.

The continuity conditions
\begin{equation}
u_1(1) - u_2(1) = 0, \quad w_1(1) - w_2(1) = 0, \quad
u'_1(1)- u'_2(1) = 0, \quad  w'_1(1)- w'_2(1) = 0
\end{equation}
for the radial wave functions and their derivatives at the boundary point $r=1$ lead to the algebraic
equations
 \begin{equation}
T_4(m,\epsilon,v.a,b,s)
\left(\begin{array}{c}
 c_{1-} \\
 c_{1+} \\
 c_{2-} \\
 c_{2+}
   \end{array}\right) =0
\end{equation}
 for coefficients  $c_{1-}, c_{1+}, c_{2-}$ and $c_{2+}$ where
 \begin{equation}
 T_4(m,\epsilon,v.a,b,s)=
  \left(\begin{array}{rrrr}
  f_{1-}(1)&f_{1+}(1)&
   -f_{2-}(1)&-f_{2+}(1) \\
   g_{1-}(1)&g_{1+}(1)&
   -g_{2-}(1)&-g_{2+}(1) \\
  f'_{1-}(1)&f'_{1+}(1)&
   -f'_{2-}(1)&-f'_{2+}(1) \\
 g'_{1-}(1)&g'_{1+}(1)&
   -g'_{2-}(1)&-g'_{2+}(1)
 \end{array}\right) .
 \end{equation}

Hence, the exact equation for energy $\epsilon(m,v,a,b,s)$ is
\begin{equation}
\det T_4(m,\epsilon,v,a,b,s) =0.
\end{equation}
This equation is solved numerically.

The desired coefficients are
\begin{equation}
\left(\begin{array}{c}
 c_{1+} \\
 c_{2-} \\
 c_{2+}
   \end{array}\right) = c_{1-} T_3^{-1}(m,\epsilon,v,a,b,s)  \left(\begin{array}{c}
 -f_{1-}(1) \\
 -g_{1-}(1) \\
 -f'_{1-}(1)
   \end{array}\right) ,
  \end{equation}
   where
\begin{equation}
   T_3(m,\epsilon,v,a,b,s) =  \left(\begin{array}{rrr}
 f_{1+}(1)&
   -f_{2-}(1)&-f_{2+}(1) \\
  g_{1+}(1)&
   -g_{2-}(1)&-g_{2+}(1) \\
  f'_{1+}(1)&
   -f'_{2-}(1)&-f'_{2+}(1)
  \end{array}\right) .
\end{equation}
 The value of $c_{1-}$ is determined by  the following normalization condition
$\int_0^{\infty}\left(u^2(r) +  w^2(r)\right)r dr = 1 $.

\section{Numerical and graphic illustrations}

Now we present some numerical and graphic illustrations in
addition to the analytical results for the ground  and first excited states in the particular cases
$m=1, m=-2$ at fixed $ s=0.05 .$

Tables show the  energies $\epsilon$ for different values of the Rashba parameter $a$,  the well depth $v$  and the magnetic field $b$ .

Figures demonstrate the examples of continuous radial wave
functions for $v=100, a=2, b=5$. Solid lines correspond to the functions $u(r)$ and
dashed lines correspond to the functions $w(r)$. We see that
   the radial wave functions  rapidly decrease outside the well.  The values of coefficients
are
$
  c_{1-} = -0.443198,  c_{1+} = 5.63821, c_{2-} = -20148.7$ and $c_{2+} =
 6147.72$
in the case of Fig. 1 and
 $
 c_{1-} = -0.065795, c_{1+} = -2.5723, c_{2-} = -31387$ and $c_{2+} =
 2251.07$
in the case of Fig. 2.

\begin{table}[b]
\caption{ Energy levels  for $m=1$.}
\begin{center}
\begin{tabular}{r r r r r| r r r r}\hline
  &$v=50$& & & &$v=100$& \\ \hline
   $b$&$a=1$& &$a=2$&  &$a=1$& &$a=2$  &\\ \hline
   $0$&$10.23$&$20.31$&$ 7.97$&$20.73$&$11.16$&$22.08$& $8.85$&$22.59$ \\
 $0.5$&$11.33$&$22.50$& $8.84$&$23.09$&$12.25$&$24.24$& $9.73$&$24.94$ \\
   $1$&$12.65$&$24.95$& $9.92$&$25.70$&$13.55$&$26.64$&$10.81$&$27.53$ \\
 $1.5$&$14.18$&$27.68$&$11.23$&$28.52$&$15.05$&$29.29$&$12.08$&$30.32$ \\
   $2$&$15.93$&$30.67$&$12.74$&$31.54$&$16.74$&$32.18$&$13.56$&$33.32$ \\
 $2.5$&$17.88$&       &$14.47$&$34.68$&$18.62$&$35.30$&$15.23$&$36.47$ \\
   $3$&       &       &$16.39$&$37.92$&$20.68$&$38.65$&$17.08$&$39.75$ \\
 $3.5$&       &       &$18.49$&       &$22.90$&$42.20$&$19.10$&$43.11$ \\
   $4$&       &       &$20.75$&       &$25.27$&$45.94$&$21.28$&$46.47$ \\
 $4.5$&       &       &$23.16$&       &$27.77$&       &$23.61$&$49.82$ \\
   $5$&       &       &$25.70$&       &       &       &$26.07$&$53.17$ \\
 $5.5$&       &       &       &       &       &       &$28.65$&$56.57$ \\
   $6$&       &       &       &       &       &       &$31.32$&$60.06$ \\
 \hline
 \end{tabular}
\end{center}
\end{table}

  \begin{table}
\caption{ Energy levels  for $m=-2$.}
\begin{center}
\begin{tabular}{r r r r r| r r r r}\hline
  &$v=50$& & & &$v=100$& \\ \hline
   $b$&$a=1$& &$a=2$&  &$a=1$& &$a=2$  &\\ \hline
   $0$&$10.23$&$20.31$&$ 7.97$&$20.73$&$11.16$&$22.08$& $8.85$&$22.59$ \\
 $0.5$& $9.36$&$18.41$& $7.33$&$18.63$&$10.26$&$20.18$& $8.18$&$20.47$ \\
   $1$& $8.71$&$16.79$& $6.89$&$16.79$& $9.57$&$18.52$& $7.70$&$18.60$ \\
 $1.5$& $8.26$&$15.45$& $6.66$&$15.21$& $9.08$&$17.12$& $7.40$&$16.98$ \\
   $2$& $8.02$&$14.37$& $6.61$&$13.91$& $8.77$&$15.96$& $7.29$&$15.59$ \\
 $2.5$& $7.96$&$13.56$& $6.74$&$12.87$& $8.63$&$15.05$& $7.33$&$14.45$ \\
   $3$& $8.08$&$12.99$& $7.01$&$12.08$& $8.66$&$14.36$& $7.53$&$13.55$ \\
 $3.5$& $8.34$&       & $7.42$&$11.53$& $8.84$&$13.90$& $7.86$&$12.87$ \\
   $4$&       &       & $7.94$&$11.21$& $9.15$&$13.64$& $8.31$&$12.41$ \\
 $4.5$&       &       & $8.55$&$11.09$& $9.58$&$13.58$& $8.86$&$12.15$ \\
   $5$&       &       & $9.24$&$11.18$&$10.11$&$13.70$& $9.50$&$12.08$ \\
 $5.5$&       &       & $9.98$&$11.43$&       &       &$10.20$&$12.19$ \\
   $6$&       &       &$10.77$&$11.84$&       &       &$10.95$&$12.46$ \\
\hline
 \end{tabular}
\end{center}
\end{table}

\vspace{2cm}
\begin{figure}[h]
    \leavevmode
 \centering\epsfig{file=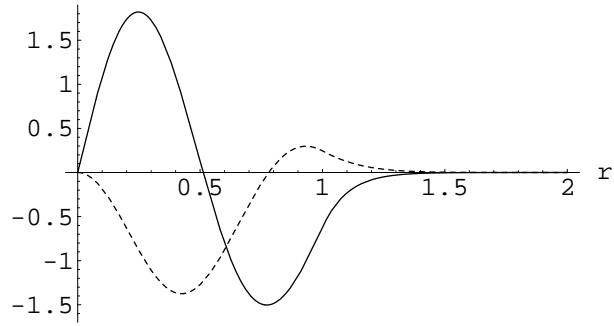}
  \caption{Radial wave functions  for $m=1, e=53.1715$.}
  \end{figure}
\begin{figure}[h!]
     \leavevmode
\centering \epsfig{file=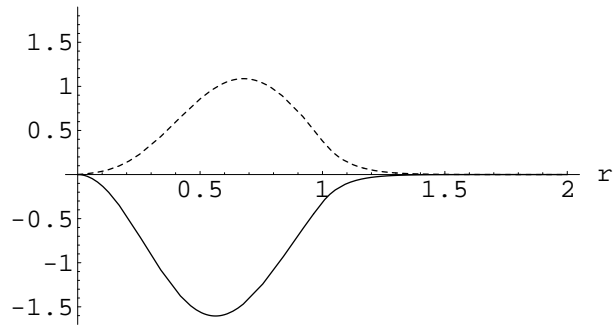} \caption{Radial wave functions   for $m=-2, e=12.0827$. }
\end{figure}

 \section{Conclusion}

So, we have constructed new exactly solvable and physically
adequate model to describe the behavior  of electron in a
semiconductor quantum dot with account of the Rashba spin-orbit
interaction and the external magnetic field.

\end{document}